\documentclass[journal,twoside]{IEEEtran}
\ifCLASSINFOpdf
   \usepackage[pdftex]{graphicx}

\else

\fi

\usepackage{lipsum}

\newcommand\blfootnote[1]{%
	\begingroup
	\renewcommand\thefootnote{}\footnote{#1}%
	\addtocounter{footnote}{-1}%
	\endgroup
}

\usepackage[cmex10]{amsmath}
\usepackage{amssymb}   
\usepackage{amsxtra}
\usepackage{amscd}
\usepackage{amsthm}

\usepackage{graphicx}   

\usepackage{array}  

\usepackage{multirow}  

\usepackage{cite}

\usepackage{booktabs}

\usepackage{color,soul} 
\soulregister\cite7
\soulregister\ref7
\soulregister\pageref7

\begin{document}
%

\begin{center}{\large \textbf{ Space-Time Channel Modulation}}\end{center}

\begin{center}Ertugrul~Basar, \textit{Senior Member, IEEE} \\ and Ibrahim Altunbas,\textit{ Member, IEEE}\end{center}

\markboth{IEEE Transactions on Vehicular Technology, Vol. PP, No. 99, February 2017}{IEEE Transactions on Vehicular Technology, Vol. PP, No. 99, February 2017}

%





\vspace*{0.4cm}
\begin{abstract}

In this paper, we introduce the concept of space-time channel modulation (STCM), which extends the classical space-time block codes into a third dimension: channel states (transmission media) dimension. Three novel STCM schemes, which provide interesting trade-offs among decoding complexity, error performance and data rate by combining space-time block coding and media-based modulation (MBM) principles, are proposed. It is shown via computer simulations that the proposed STCM schemes achieve considerably better error performance than the existing MBM and classical systems. \blfootnote{Copyright (c) 2017 IEEE. Personal use of this material is permitted. However, permission to use this material for any other purposes must be obtained from the IEEE by sending a request to pubs-permissions@ieee.org.\\
\hspace*{0.3cm}Manuscript received June 17, 2016; revised January 6, 2017; accepted February 22, 2017. Date of publication XXX, 2017;
date of current version XXX, 2017. The review of this paper was coordinated by Prof. Y. L. Guan.\\
\hspace*{0.3cm}The authors are with Istanbul Technical University, Faculty of Electrical and Electronics Engineering, 34469, Istanbul, Turkey. (e-mail: basarer@itu.edu.tr; ibraltunbas@itu.edu.tr).}  
  
\end{abstract}
\begin{IEEEkeywords}
index modulation, media-based modulation, space-time block coding, space shift keying,  ML detection.
\end{IEEEkeywords}

%
\IEEEpeerreviewmaketitle


\section{Introduction}

\IEEEPARstart{I}{ndex} modulation (IM) techniques have attracted significant attention in the past few years due to their advantages such as improved energy/spectral efficiency and error performance over classical digital modulation schemes. In an IM scheme, additional information bits can be transmitted by means of the indices of the building blocks of the target communication systems \cite{IM_5G}. Two well-known applications of the IM concept are space shift keying/spatial modulation (SSK/SM) \cite{SM_jour,SSK,Design_SM} and orthogonal frequency division multiplexing with index modulation (OFDM-IM) \cite{OFDM_IM} schemes, in which the indices of the available transmit antennas and OFDM subcarriers are considered for the transmission of additional information bits, respectively. SM and OFDM-IM systems have been extensively studied in the past few years and the researchers have investigated interesting design issues. In\cite{SM_Survey_2016}, the potential of SM systems has been explored for frequency-selective fading channels, and single-carrier SM schemes with cyclic prefix have been considered as promising alternatives for low complexity broadband multiple-input multiple-output (MIMO) systems. As a candidate for next-generation wireless communication systems, OFDM-IM has been combined with MIMO transmission in\cite{MIMO_OFDM_IM_2} and improvements have been reported over classical MIMO-OFDM systems.   

\textit{Media based modulation (MBM)} introduced by Khandani in \cite{Khandani1,Khandani2,Khandani3}, is a novel IM scheme in which the concept of reconfigurable antennas (RAs) is used in a clever fashion for the transmission of additional information bits. In an MBM scheme, the characteristics of transmit antennas (equipped with either radio frequency (RF) mirrors \cite{Khandani3} or electronic switches \cite{RA_SSK}) are changed according to the information bits unlike the conventional RA schemes. In other words, MBM considers a finite number of channel states and the information is embedded into the determination of the active channel state. In this context, MBM performs the modulation of the wireless channel itself; therefore, can be considered as \textit{channel modulation}. The concept of MBM is introduced in \cite{Khandani1} with the aim of embedding information to the variations of channel states (transmission media). It has been shown in \cite{Khandani2} that significant gains can be obtained by MBM over classical single-input multiple-output (SIMO) systems and MBM can be practically realized using RF mirrors. In \cite{Khandani3}, MBM is adapted to MIMO transmission to obtain a more flexible system design with reduced complexity implementation. SSK and MBM principles are combined in \cite{RA_SSK} to improve the error performance of SSK considering correlated and nonidentically distributed  Rician fading channels. Later, RA-based SSK\cite{RA_SSK} is considered for underlay cognitive radio systems in Rician fading channels and improvements are shown compared to conventional spectrum sharing systems\cite{RA_SSK2}. More recently, MBM and generalized SM techniques are combined in \cite{MBM_GSM} and \cite{MBM_TVT} to improve the error performance of MIMO-MBM and an Euclidean distance-based RF mirror activation pattern selection procedure is proposed.

Against this background, in this paper, we introduce the concept of \textit{space-time channel modulation (STCM)}, by exploiting not only the space and time domains to achieve transmit diversity but also the channel states domain to convey additional information bits. In the proposed STCM scheme, the incoming information bits determine the wireless channel states to be used as well as the complex data symbols to be transmitted using space-time block coding/codes (STBC(s)). Three novel STCM schemes, which offer interesting trade-offs among decoding complexity, error performance and data rate, are proposed and the reduced complexity maximum likelihood (ML) detector of the STCM scheme is formulated. Finally, the theoretical average bit error probability (ABEP) upper bound of STCM is derived and the superiority of STCM schemes over existing methods are shown via extensive computer simulations.


\vspace*{-0.05cm}

\section{The Channel Modulation: MBM and SSK}
MBM is a novel digital modulation scheme, in which the information bits are mapped to the indices of the available channel states (by the adjustment of on/off status of the available RF mirrors or changing the characteristics of the transmit antennas) in addition to the classical two dimensional signal constellations. In classical $Q$-ary modulation schemes, such as $Q$-QAM/PSK, the amplitude and/or phase of  a carrier signal are adjusted according to the information bits. On the other hand, MBM aims to convey information bits by not only changing the parameters of a carrier signal but also the variation of the wireless channel itself as long as the receiver has the knowledge of the possible transmission scenarios. In other words, MBM is capable of altering the channel fading coefficients for each combination of the active RF mirrors (for each channel state). In the following, we present the conceptual similarities between SSK and MBM schemes.

Consider the operation of the SSK scheme for a flat Rayleigh fading $T\times R $ MIMO system, where $T$ and $R$ denote the number of transmit and receive antennas, respectively. For each transmission interval, according to the SSK principle, only one transmit antenna, whose index is given by $t$, where $t \in \left\lbrace 1,2,\ldots,T\right\rbrace $, is activated and the received signals can be expressed as $\mathbf{y}=\mathbf{Hx}+\mathbf{n} = \mathbf{h}_t + \mathbf{n} $, where $ \mathbf{y} \in \mathbb{C}^{R \times 1} $ is the vector of received signals, $ \mathbf{H} \in \mathbb{C}^{R \times T} $ is the MIMO channel matrix, whose elements are independent and identically distributed (i.i.d.) with $\mathcal{CN}(0,1)$ distribution, where $\mathcal{CN}(0,\sigma^2)$ represents circularly symmetric complex Gaussian distribution with variance $\sigma^2$. $\mathbf{h}_t$ is the $t$th column of $ \mathbf{H} $. $ \mathbf{x} \in \mathbb{R}^{T \times 1} $ is the corresponding transmission vector of SSK whose $t$th element is non-zero only (i.e., $1$ in the baseband),  and $ \mathbf{n} \in \mathbb{C}^{R \times 1} $ is the vector of additive white Gaussian noise samples whose elements follow $\mathcal{CN}(0,N_0)$ distribution.

On the other hand, let us consider the employment of an MBM scheme (without transmitting additional bits with $Q$-ary constellations, i.e., using a carrier with constant parameters) for a $1\times R$ SIMO system with $M$ available RF mirrors at the transmitter. A total of $2^{M}$ channel states can be obtained by adjusting the on/off status of the available RF mirrors according to $M $ information bits \cite{Khandani3}. As an example, two bits determine the on/off status of the RF mirrors and four different channel realizations can be obtained for $M=2$ as seen from Table I. Let us denote the corresponding vector of channel fading coefficients between the transmit antenna and receive antennas for channel state $i$ by $\mathbf{h}_i \in \mathbb{C}^{R \times 1}$, where $i=1,2,\ldots, 2^{M}$. Similar to SSK, the information is conveyed over the channel realization itself in the MBM scheme and for each transmission interval, one channel state $i$ is selected for which the received signals can be expressed as
\begin{align}
\mathbf{y}=\mathbf{Gz}+\mathbf{n}=\mathbf{h}_i + \mathbf{n}
\label{2}
\end{align}
where $ \mathbf{y}  $ and $ \mathbf{n}  $ are the same as defined for SSK, $ \mathbf{G}  = \begin{bmatrix}
\mathbf{h}_1 & \mathbf{h}_2 & \cdots & \mathbf{h}_{2^{M}}
\end{bmatrix} \in \mathbb{C}^{R \times 2^{M}} $ is the corresponding extended channel matrix of MBM, which contains all possible $2^{M}$ realizations of the SIMO channel and assumed to be consisted of i.i.d. elements with $\mathcal{CN}(0,1)$ distribution\footnote{In this paper, we assume that there is no correlation between the channel coefficients of different channel states. In the presence of correlated channel states, a degradation can be expected in the attainable error performance; however, since SM/SSK and MBM systems are analogous, efficient solutions from the SM literature can be adapted for MBM based schemes. Interested readers are referred to\cite{Design_SM},\cite{SM_Survey_2016} and the references therein.  }. $ \mathbf{z} \in \mathbb{R}^{2^{M} \times 1} $ is the transmission vector of MBM whose $i$th element is non-zero only. ML detection of MBM can be easily performed by
\begin{equation}
\mathbf{\hat{z}}=\arg \min_{\mathbf{z}} \left\| \mathbf{y}- \mathbf{Gz} \right\|^2. 
\end{equation}

\begin{table}[t]
	\centering
	\setlength{\extrarowheight}{1pt}
	\caption{MBM for a $1\times R$ SIMO System with two RF mirrors}
	\vspace*{-0.2cm}
	\label{my-label}
	\begin{tabular}{cccc} \hline \hline
		Bits & \begin{tabular}[c]{@{}c@{}}Status of\\ RF Mirrors\end{tabular} & \begin{tabular}[c]{@{}c@{}}Active channel\\ state index ($ i $) \end{tabular} & \begin{tabular}[c]{@{}c@{}}Transmission\\ vector $(\mathbf{z}^\textrm{T})$\end{tabular} \\ 	 \midrule 
		$ \left\lbrace 0,0 \right\rbrace  $	& $1\textrm{st}\rightarrow\textrm{off}$, $2\textrm{nd}\rightarrow\textrm{off}$ &  $ 1 $ &  $ \begin{bmatrix}
		1 & 0 & 0 & 0 \end{bmatrix}  $ \\
		$ \left\lbrace 0,1 \right\rbrace  $	 & $1\textrm{st}\rightarrow\textrm{off}$, $2\textrm{nd}\rightarrow\textrm{on}$	& $ 2 $ & $ \begin{bmatrix}
		0 & 1 & 0 & 0 \end{bmatrix} $  \\
		$ \left\lbrace 1,0 \right\rbrace  $	& $1\textrm{st}\rightarrow\textrm{on}$, $2\textrm{nd}\rightarrow\textrm{off}$	& $ 3 $ & $ \begin{bmatrix}
		0 & 0 & 1 & 0 \end{bmatrix} $  \\
		$ \left\lbrace 1,1\right\rbrace  $	& $1\textrm{st}\rightarrow\textrm{on}$, $2\textrm{nd}\rightarrow\textrm{on}$	& $ 4 $ & $ \begin{bmatrix}
		0 & 0 & 0 & 1 \end{bmatrix} $ \\ \hline \hline
	\end{tabular}
	\vspace*{-0.4cm}
\end{table}

\textit{Remark 1}: Comparing the signal models of SSK and MBM, we observe that SSK and MBM schemes operate in a very similar fashion by transmitting information bits using different realizations of the wireless channels. We conclude that assuming $R$ receive antennas, SSK with $T$ transmit antennas and MBM with $M$ RF mirrors (using a single transmit antenna) are \textit{identical} for $T=2^{M}$. In other words, $M=\log_2(T)$ bits can be transmitted by using either the SSK scheme with a $T\times R$ MIMO system or the MBM scheme with a $1\times R$ system and $M$ RF mirrors. However, MBM cleverly overcomes the main limitation of the SSK scheme by linearly increasing the number of transmitted information bits by the number of RF mirrors. As an example, to achieve a data rate of $\eta=10$ bits per channel use (bpcu), SSK requires $2^{10}$ transmit antennas, while MBM can handle the same transmission using only a single transmit antenna equipped with $10$ RF mirrors.

Using the signal model of (\ref{2}), the ABEP of the MBM scheme can be easily obtained by the calculation of conditional pairwise error probability (CPEP) for the erroneous detection of $\mathbf{z}$ to $\mathbf{\hat{z}}$ as follows:
\begin{equation}
P\left( \mathbf{z} \rightarrow  \mathbf{\hat{z}} \!\left. \right|\! \mathbf{G} \right) \!=\! P\left( \left\| \mathbf{y}- \mathbf{G } \mathbf{\hat{z}} \right\|^2 \!\!<\! \left\| \mathbf{y}- \mathbf{Gz} \right\|^2 \right)\!=\! Q\left( \! \sqrt{\frac{\Gamma }{2N_0}} \right)  
\end{equation}
where $\Gamma=\left\|\mathbf{G}(\mathbf{z}-\mathbf{\hat{z}}) \right\|^2$. Considering the quadratic form of $\Gamma=\sum\nolimits_{r=1}^{R} \mathbf{G}_{r*} \mathbf{Q} \mathbf{G}_{r*}^{\textrm{H}}$ where $\mathbf{Q}=\left(\mathbf{z}-\mathbf{\hat{z}} \right) \left(\mathbf{z}-\mathbf{\hat{z}} \right)^{\textrm{H}} $ and $\mathbf{G}_{r*}$ is the $r$th row of $\mathbf{G}$, since $ \mathbf{G}_{r*} $'s are i.i.d. for all $r$ and $\textrm{rank} \left(\mathbf{Q }\right)=1 $, the moment generating function (MGF) of $\Gamma$ is obtained as \cite{Turin} $M_{\Gamma}(s)=\big(1-s \left\|(\mathbf{z}-\mathbf{\hat{z}}) \right\|^2 \big)^{-R} $, which yields the following unconditional PEP (UPEP) using the alternative form of the $ Q $-function:
\\\vspace*{-0.25cm}
\begin{equation}
P\left( \mathbf{z} \rightarrow  \mathbf{\hat{z}} \right)=\frac{1}{\pi} \int_{0}^{\pi/2}\Bigg( \frac{\sin^2 \theta}{\sin^2\theta + \frac{\left\|\mathbf{z}-\mathbf{\hat{z}} \right\|^2}{4N_0} }\Bigg) ^R d\theta
\end{equation}
which can be calculated from \cite{Simon}, Eq. (5A.4a). Due to the special form of the MBM transmission vectors, we obtain $ \left\|\mathbf{z}-\mathbf{\hat{z}} \right\|^2=2 $ for all $\mathbf{z}$ and $\mathbf{\hat{z}}$ when $\mathbf{z} \neq \mathbf{\hat{z}} $. Considering the symmetry of transmission vectors, the ABEP upper bound of MBM is obtained as $  P_b \le \frac{1}{M} \sum\nolimits_{\mathbf{\hat{z}}} P\left( \mathbf{z} \rightarrow  \mathbf{\hat{z}} \right) n\left( \mathbf{z},\mathbf{\hat{z}}\right)  $  where without loss of generality, $\mathbf{z}$ can be selected as $\mathbf{z}=\begin{bmatrix}
1 & 0 & \cdots & 0
\end{bmatrix}^{\textrm{T}}$ and $ n\left( \mathbf{z},\mathbf{\hat{z}}\right) $ denotes the number of bit errors for the corresponding pairwise error event. The above analysis can be easily extended to $Q$-QAM/PSK aided (using a carrier with varying parameters) MBM with $\eta=M+\log_2(Q)$ bpcu, which is analogous to SM.

\textit{Remark 2}: Since SSK and MBM schemes are identical for the same data rate $(T={{2}^{M}})$, they have exactly the same BER performance for the same number of receive antennas $(R)$ and the above analysis is also valid for SSK.

\begin{figure}[!t]
	\begin{center}\resizebox*{\columnwidth}{7cm}{\includegraphics{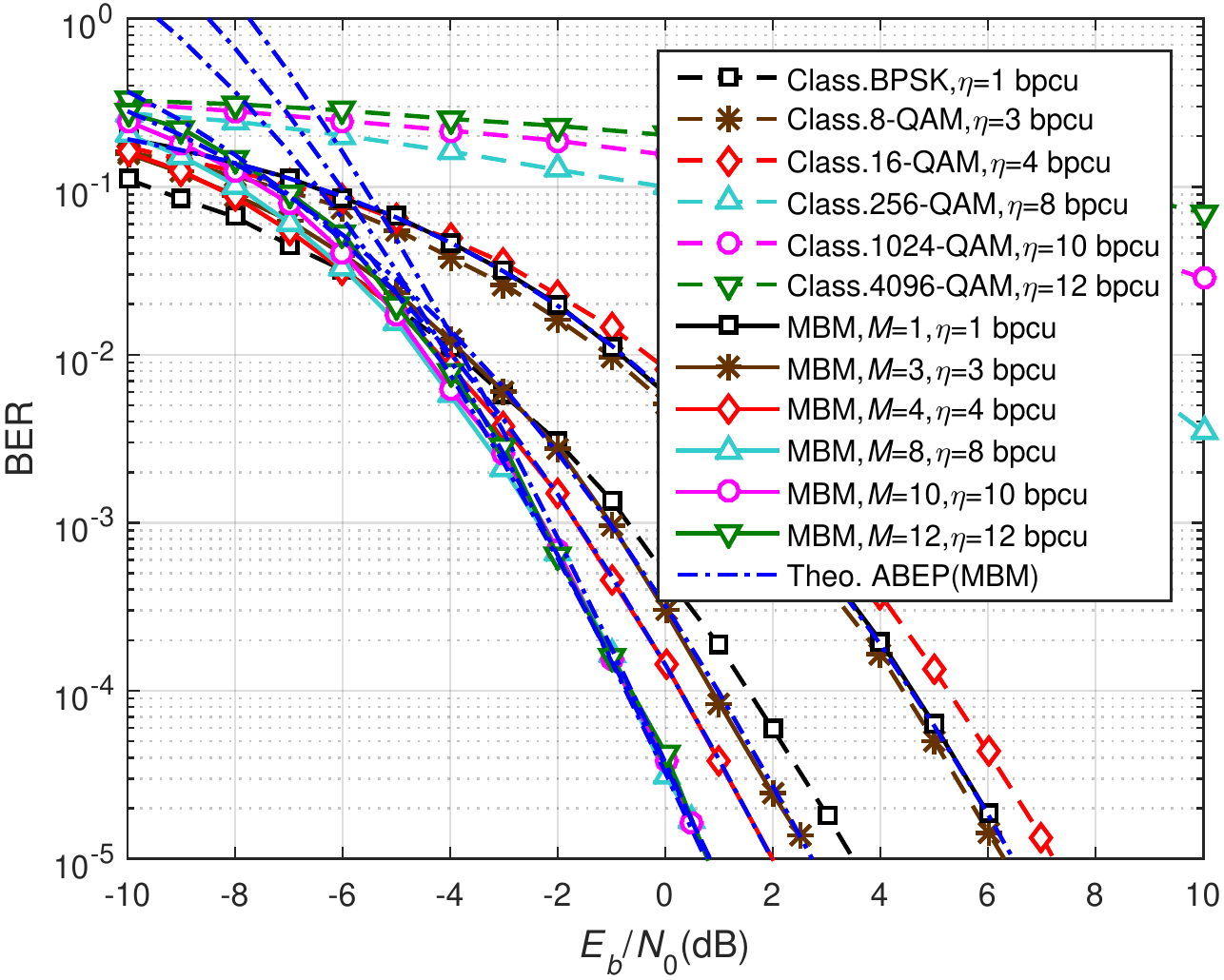}}
		\vspace*{-0.7cm}\caption{BER performance comparison of classical SIMO and MBM-SIMO schemes for different data rates, $1\times 8$ SIMO system with theoretical ABEP curves.}\vspace*{-0.5cm}
	\end{center}
\end{figure}

In Fig. 1, we compare the bit error rate (BER) performance of MBM and classical SIMO schemes for different data rates and $R=8$, where the derived theoretical ABEP curves are shown with dash-dot lines. As seen from Fig. 1, MBM provides significant improvements in required signal-to-noise ratio in terms of $E_b/N_0$ compared to classical SIMO systems, where $E_b$ is the average transmitted energy per bit. This improvement can be explained by the fact that for higher data rates, classical SIMO scheme requires higher order constellations, which are composed of more closely spaced elements; on the other hand, the MBM scheme can increase the data rate while keeping the same distance between the transmission vectors similar to the trend observed in frequency shift keying modulation. It is also interesting to note that a close BER performance is obtained for MBM schemes with $M\ge 8$ due to the definition of the SNR as ${{E}_{b}}/{{N}_{0}}$, where ${{E}_{b}}=1/M$.

\begin{figure}[!t]
	\begin{center}\resizebox*{\columnwidth}{7cm}{\includegraphics{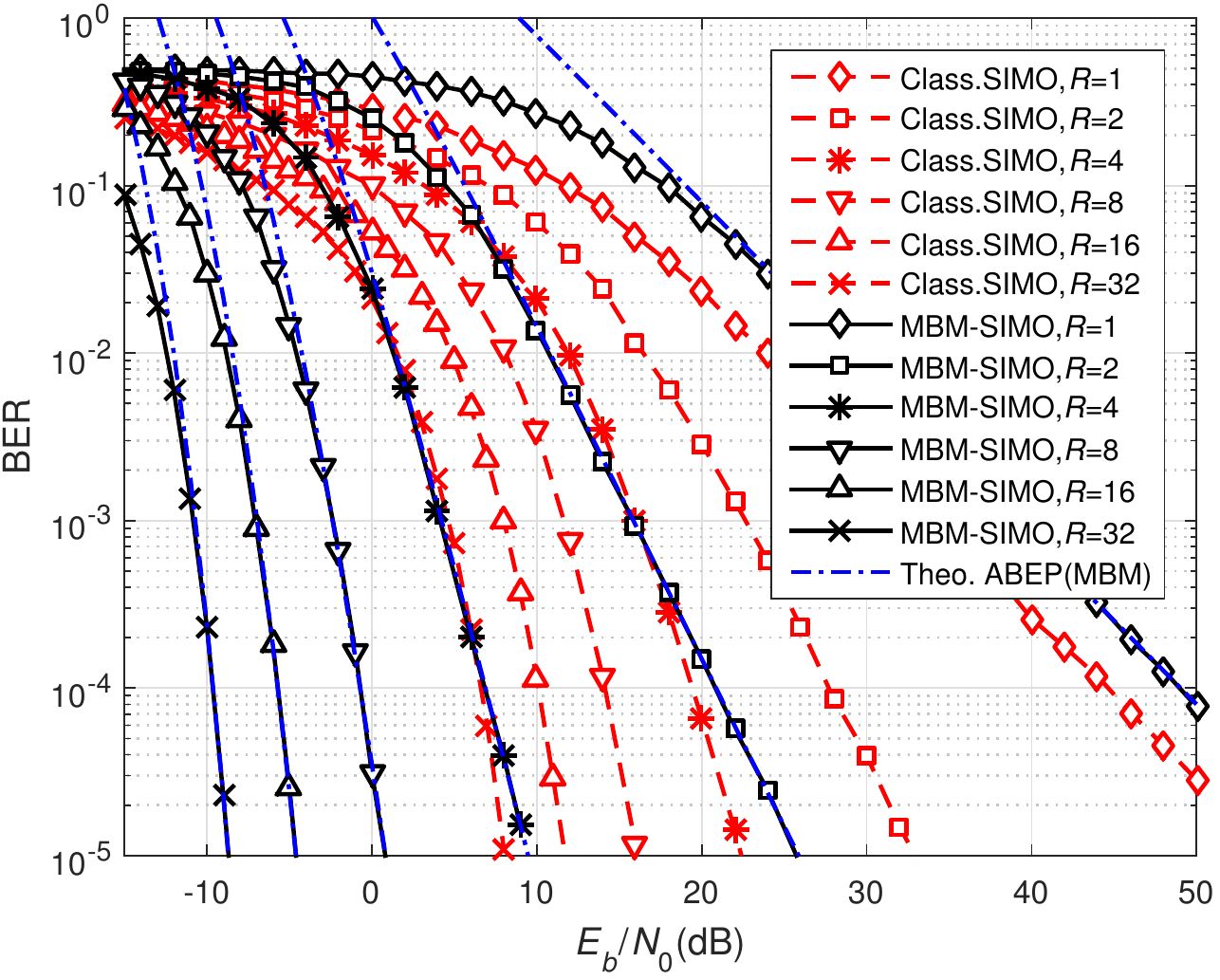}}
		\vspace*{-0.7cm}\caption{BER performance comparison of classical SIMO and MBM-SIMO schemes for different number of receive antennas, $\eta=8$ bpcu transmission (class. SIMO with 256-QAM, MBM-SIMO with $M=8$ RF mirrors).}\vspace*{-0.7cm}
	\end{center}
\end{figure}

In Fig. 2, we investigate the effect of increasing number of receive antennas for MBM and classical SIMO schemes, where we considered $8$ bpcu transmission for all cases. As seen from Fig. 2, MBM scheme is outperformed by the classical scheme for only $R=1$, i.e., for the SISO case; however, it provides significantly better BER performance in all other cases since it benefits more from increasing number of receive antennas due to the transmission of the data with channel realizations.

\section{Space-Time Channel Modulation}
In this section, we introduce the concept of STCM by extending the classical space-time codes into a new third dimension, which is the channel states (media) dimension. As the core STBC, we consider the Alamouti's STBC given by
\begin{equation}
\mathbf{S}=\begin{bmatrix}
x_1 & -x_2^* \\
x_2 & x_1^*
\end{bmatrix}
\label{STBC}
\end{equation}
where the rows and columns stand for transmit antennas and time slots, respectively, $x_1,x_2 \in \mathcal{S}$ and $ \mathcal{S} $ denotes $Q$-ary signal constellation. The classical Alamouti's STBC can transmit two complex symbols, i.e., $2\log_2(Q)$ bits, in two time slots for which we obtain $\eta=\log_2(Q)$ bpcu. By extending $\mathbf{S}$ into channel states, we aim to improve the data rate while ensuring transmit diversity and/or simplified ML detection.

\begin{figure*}[t]
	\begin{center}
		{\includegraphics[scale=0.72]{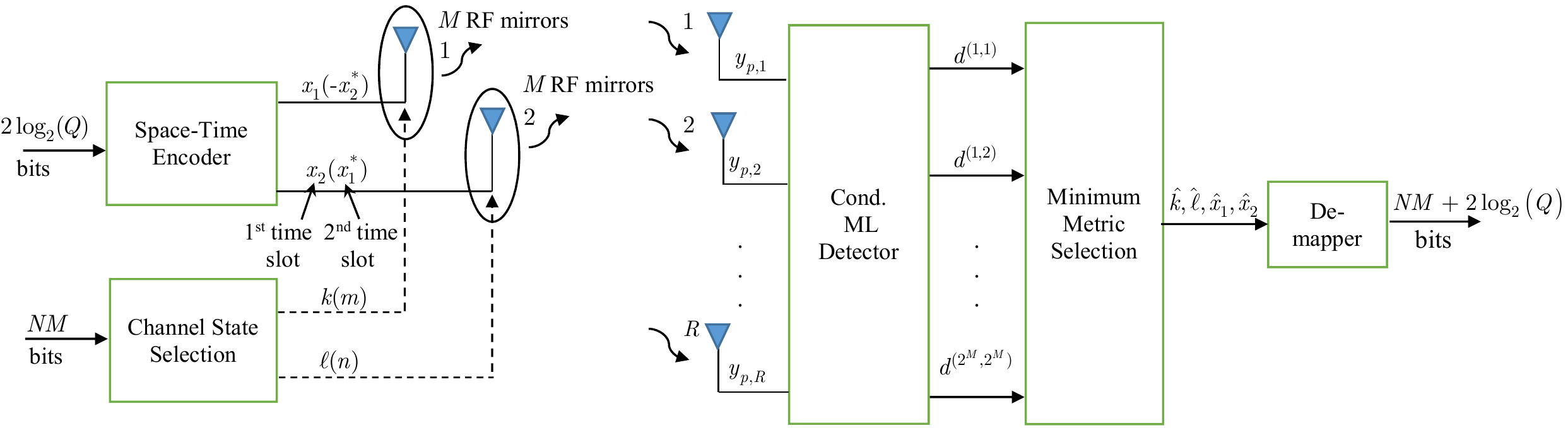}}
		\vspace*{-0.6cm}
		\caption{Transceiver structure of the STCM scheme for a $2\times R$ MIMO system ($N=1$ for Scheme 2 and $N=2$ for Schemes 1 and 3)}
		\vspace*{-0.4cm}
	\end{center}
\end{figure*}

The block diagram of the STCM transceiver is shown in Fig. 3. In STCM, additional information bits can be transmitted by changing the channel state for each transmit antenna in each time slot. Assume that we want to transmit the STBC of (\ref{STBC}) in two time slots using two transmit antennas, where we can determine the channel state for each transmit antenna using $M$ RF mirrors at each of them. As an example, combining MBM principle with STBC, we can transmit $\mathbf{S}$ by selecting channel states $k$ and $l$ for the first and second transmit antennas, respectively, where $k,l \in \left\lbrace 1,2,\ldots,2^{M}\right\rbrace $ and assume that we select the same channel states in the second time slot. Therefore, a total of $2M+ 2\log_2(Q) $ bits can be transmitted by this generic STCM scheme in two time slots, which leads to an average of $\eta=M+\log_2(Q) $ bpcu transmission. 

The baseband transmission model for the classical STBC of (\ref{STBC}) can be given as $ \mathbf{Y}=\mathbf{H} \mathbf{S} + \mathbf{N} $, where $ \mathbf{Y} \in \mathbb{C}^{R \times 2} $ is the matrix of received signals in two time slots, $ \mathbf{H} \in \mathbb{C}^{R \times 2} $ is the matrix of MIMO channel fading coefficients and $ \mathbf{N} \in \mathbb{C}^{R \times 2} $ is the matrix of noise samples. For the STCM scheme, the extension of $\mathbf{S}$ into channel states can be represented by the following signal model:
\begin{equation}
\mathbf{Y}=\mathbf{C} \mathbf{Z} + \mathbf{N}
\label{10}
\end{equation}    
where $ \mathbf{Y} $ and $ \mathbf{N} $ are the same as defined earlier, while $\mathbf{C}=\begin{bmatrix}
\mathbf{h}_1 & \cdots & \mathbf{h}_{2^{M}} & \mathbf{h}_{2^{M} +1 } & \cdots & \mathbf{h}_{ 2^{(M+1)}}  
\end{bmatrix} \in \mathbb{C}^{R \times 2^{(M+1)}}$ 
is the extended STCM channel matrix, which considers all possible channel state realizations for both transmit antennas, where $\mathbf{h}_k\in \mathbb{C}^{R\times 1}$ and $\mathbf{h}_{2^M+l}\in \mathbb{C}^{R\times 1}$ for $k,l \in \left\lbrace 1, 2,\ldots,2^{M}\right\rbrace  $ denote the vector of channel coefficients between the receive antennas and the first and second transmit antennas for channel states $k$ and $l$, respectively. We assume that the elements of $ \mathbf{C} $ and $ \mathbf{N} $ follow $ \mathcal{CN}(0,1) $ and $ \mathcal{CN}(0,N_0) $ distributions, respectively. On the other hand, $\mathbf{Z} \in \mathbb{C}^{ 2^{(M+1)}\times 2} $ is the extended version of $\mathbf{S}$, which has the following general form:
\\ \vspace*{-0.6cm}
 \begin{align}
 &\hspace*{1.95cm} \downarrow^{k} \hspace*{3cm} \downarrow^{2^{M}+l}  \nonumber \\
 &\mathbf{Z}=\left[
 \begin{array}{cccccc|cccccccc} 	 	
 0 & \cdots    & x_1  &  \cdots  & &  0 & 0   & \cdots    & x_2  &  \cdots  & &  0 \\ 
 0 & \cdots &   & -x_2^*  & \cdots  & 0  &  0 & \cdots &   & x_1^*  & \cdots  & 0 
 \end{array} \right] ^{\textrm{T}} \nonumber 	\\
 &\hspace*{2.7cm} \uparrow_{m} \hspace*{2.75cm} \uparrow_{2^{M}+n}  
 \label{Z}  	
 \end{align} 
\vspace*{-0.6cm}\\
where we assume that in the first time slot, channel states $k$ and $l$ are selected for the transmission of $x_1$ and $x_2$ from the first and second transmit antennas, respectively. On the other hand, in the second time slot, channel states $m$ and $n$ are selected for the transmission of $-x_2^*$ and $x_1^*$ from the first and second transmit antennas, respectively. As seen from (\ref{Z}), the STCM scheme can be considered as a dynamic STBC, similar to the space-time block coded SM (STBC-SM) scheme \cite{STBC_SM}, in which the transmission matrices are obtained by extending $\mathbf{S}$ into antenna (spatial) domain. On the other hand, STBC-SM scheme requires more than two transmit antennas to map additional information bits, while the STCM scheme can benefit from RF mirrors and transmit additional number of bits by using only two transmit antennas as seen from Fig. 3. However, both STCM and STBC-SM schemes suffer from a loss in data rate due to the use of a second time slot, which is a necessity to achieve additional transmit diversity.    

Considering the generic form of (\ref{Z}), in the following, we propose three different STCM schemes, which provide interesting trade-offs among decoding complexity, data rate and error performance.

For the transmission of two complex symbols ($x_1$ and $x_2$) and their conjugates ($-x_2^*$ and $x_1^*$), three novel STCM schemes are proposed, which have the following parameters:
\begin{align}
&\text{Scheme 1:}\quad m=k,\, n=l,\, \eta=M+ \log_2(Q) \,\, \text{bpcu} \nonumber \\
 &\text{Scheme 2:}\quad k=l=m=n,\, \eta=0.5M+\log_2(Q) \,\, \text{bpcu} \nonumber \\
 &\text{Scheme 3:}\quad m=l,\, n=k,\, \eta=M+ \log_2(Q) \,\, \text{bpcu}.
 \label{eq:8}
\end{align}
Scheme 1 is the generic STCM scheme described earlier. For Scheme 2, the same channel state $(k)$ is selected for both transmit antennas and both time slots, while for Scheme 3, in the first time slot, channel states $k$ and $l$ are selected for the first and second transmit antennas, respectively; however, for the second time slot, channel states $l$ and $k$ are selected for the first and second transmit antennas, respectively.

To perform reduced complexity ML detection, for a given quadruplet $(k,l,m,n)$, the following equivalent signal model can be obtained from (\ref{10}): $ \mathbf{y}_{eq}=\mathbf{C}_{eq} \mathbf{z}_{eq} + \mathbf{n}_{eq} $, where $\mathbf{y}_{eq}=\begin{bmatrix}
y_{1,1} & y_{2,1}^* & y_{1,2} & y_{2,2}^* & \cdots & y_{1,R} & y_{2,R}^* &
\end{bmatrix}^{\textrm{T}}\in \mathbb{C}^{2R\times 1}$ is the equivalent received signals vector and $y_{p,r}$ is the received signal at receive antenna $r$ at time slot $p$, $ \mathbf{z}_{eq}=\begin{bmatrix}
x_1 & x_2
\end{bmatrix}^{\textrm{T}} $ and $\mathbf{n}_{eq}\in \mathbb{C}^{2R\times 1}$ represent equivalent data symbols and noise vectors, respectively. $\mathbf{C}_{eq} \in \mathbb{C}^{2R\times 2}$ is the equivalent STCM channel matrix, which has the following general form:
\begin{equation}
\mathbf{C}_{eq}\!=\!\begin{bmatrix}
\mathbf{c}_1 & \!\mathbf{c}_2
\end{bmatrix}\!=\! 
\begin{bmatrix}
h_{k,1} &  h_{2^M+n,1}^* &\cdots &  h_{k,R} & h_{2^M+n,R}^* \\
h_{2^M+l,1} & -h_{m,1}^* & \cdots & h_{2^M+l,R} & -h_{m,R}^* 
\end{bmatrix}^\textrm{\!T}
\label{C_eq}
\end{equation}
where $h_{k,r}$ and $h_{m,r}$ denote the channel fading coefficient between the first transmit antenna and $r$th receive antenna for channel states $k$ and $m$, respectively, while $h_{2^M+l,r}$ and $h_{2^M+n,r}$ denote the channel fading coefficient between the second transmit antenna and $r$th receive antenna for channel states $l$ and $n$, respectively and $\mathbf{h}_i=\begin{bmatrix}
h_{i,1} & h_{i,2} & \cdots & h_{i,R}
\end{bmatrix}^{\textrm{T}}$ for $i=1,2,\ldots,2^{(M+1)}$. Brute-force ML detection can be performed for Scheme 1 (or 3) and Scheme 2 by 
\begin{align}
\big(\hat{k},\hat{l},\hat{x}_1,\hat{x}_2 \big)&=\arg \min_{k,l,x_1,x_2}\left\| \mathbf{y}_{eq}-\mathbf{C}_{eq} \mathbf{z}_{eq}\right\| ^2 \nonumber \\
\big(\hat{k},\hat{x}_1,\hat{x}_2 \big)&=\arg \min_{k,x_1,x_2}\left\| \mathbf{y}_{eq}-\mathbf{C}_{eq} \mathbf{z}_{eq}\right\| ^2 
\end{align}
which require $ 2^{2M} Q^2 $ and $2^{M} Q^2 $ metric calculations, respectively. However, as seen from (\ref{C_eq}), for Schemes 1 and 2, the clever selection of the active channel states ensures that $\mathbf{c}_1^{\textrm{H}}\mathbf{c}_2=0$ for all possible realizations of $ \mathbf{C}_{eq} $, which allows the independent detection of $x_1$ and $x_2$. In what follows, we present the reduced complexity conditional ML detection of Scheme 1, while the same procedures can be followed for Scheme 2 by considering $k=l$. 

For Scheme 1, considering a given pair of $(k,l)$, the minimum ML decision metrics can be obtained by the conditional STCM ML detector for $x_1$ and $x_2$, respectively as $ m_1^{(k,l)}=\min_{x_1} \left\|\mathbf{y}_{eq}-\mathbf{c}_1 x_1 \right\|^2 $ and $ m_2^{(k,l)}=\min_{x_2} \left\|\mathbf{y}_{eq}-\mathbf{c}_2 x_2 \right\|^2  $, and the corresponding conditional ML estimates of $x_1$ and $x_2$ are obtained as follows: $ x_{1}^{(k,l)}=\arg\min_{x_1} \left\|\mathbf{y}_{eq}-\mathbf{c}_1 x_1 \right\|^2 $ and $ x_{2}^{(k,l)}=\arg\min_{x_2} \left\|\mathbf{y}_{eq}-\mathbf{c}_2 x_2 \right\|^2 $. Afterwards, the minimum ML decision metric for a given pair of $(k,l)$  is obtained as $d^{(k,l)}=m_1^{(k,l)} + m_2^{(k,l)}$. After the calculation of $d^{(k,l)}$ for all possible pairs of $(k,l)$, the minimum metric selector determines the most likely combination of the active channel states by $ (\hat{k},\hat{l} ) = \arg \min_{(k,l)} d^{(k,l)}  $ and the corresponding data symbols from $ (\hat{x}_1,\hat{x}_2 ) =  (x_1^{(\hat{k},\hat{l})},x_2^{(\hat{k},\hat{l})} ) $. Finally, the detection of the input bits can be performed by bit demapping operation. Therefore, using conditional ML detection, the total number of required metric calculations is reduced from $2^{2M} Q^2  $ to $ 2^{2M+1} Q $ for Scheme 1. Similarly for Scheme 2, the total number of required metric calculations can be reduced from $2^{M} Q^2  $ to $ 2^{M+1} Q $. On the other hand, since Scheme 3 does not satisfy the orthogonality of $\mathbf{C}_{eq}$, its ML detection complexity remains at $2^{2M} Q^2  $.

\section{Performance Analysis of STCM} 
In this section, we evaluate the theoretical ABEP of the STCM scheme. Considering the system model of (\ref{10}) and dynamic structure of the STCM transmission matrices, if $\mathbf{Z}$ is transmitted and it is erroneously detected as $\mathbf{\hat{Z}}$, the well-known CPEP expression can be given as \cite{Jafarkhani}
\begin{equation}
P\left(\mathbf{Z} \rightarrow \mathbf{\hat{Z}}\left. \right| \mathbf{C}  \right) = Q\left(\sqrt{\Delta / (2N_0)} \right) 
\end{equation}
where $\Delta=\big\| \mathbf{C}\big(\mathbf{Z} - \mathbf{\hat{Z}} \big) \big\|_{\textrm{F}}^2 $. Considering the quadratic form of $\Delta = \sum\nolimits_{r=1}^{R} \mathbf{C}_{r*}\mathbf{Q} \mathbf{C}_{r*}^{\textrm{H}} $ and i.i.d. elements of $ \mathbf{C}_{r*} $, i.e., $E\left\lbrace \mathbf{C}_{r*}^{\textrm{H}}  \mathbf{C}_{r*}  \right\rbrace = \mathbf{I}_{2^{(M+1)}} $, we obtain the MGF of $\Delta$ as $M_{\Delta}(s)=\left[ \mathrm{det} \left( \mathbf{I}_{2^{(M+1)}} - s\mathbf{Q}  \right)\right] ^{-R} = \prod_{d=1}^{D}\left(1-s \lambda_d \right)^{-R}  $, where $\mathbf{Q}=\big(\mathbf{Z} - \mathbf{\hat{Z}} \big)\big(\mathbf{Z} - \mathbf{\hat{Z}} \big)^{\textrm{H}}$, $ D=\mathrm{rank}(\mathbf{Q}) $ and $\lambda_d,d \in \left\lbrace 1,2 \right\rbrace $ are the non-zero eigenvalues of $\mathbf{Q}$. Consequently, using the alternative form of the $Q$-function, the UPEP of the STCM scheme can be obtained as
\begin{equation}
P\big(\mathbf{Z} \rightarrow \mathbf{\hat{Z}}  \big) = \frac{1}{\pi} \int_{0}^{\pi/2} \prod\nolimits_{d=1}^{D} \left(1 + \frac{\lambda_d}{4 N_0 \sin^2\theta} \right)^{-R} d\theta 
\label{UPEP}
\end{equation}
which has a closed-form solution in \cite{Simon}, Eq. (5A.74). The ABEP upper bound of the STCM scheme is obtained as $ P_b \le \frac{1}{2^{2\eta}} \sum\nolimits_{\mathbf{Z}}^{} \sum\nolimits_{\mathbf{\hat{Z}}}^{} \frac{P(\mathbf{Z} \rightarrow \mathbf{\hat{Z}}  ) e(\mathbf{Z} , \mathbf{\hat{Z}}  )}{2\eta}  $,
where $ e(\mathbf{Z} , \mathbf{\hat{Z}}  ) $ denotes the number of bit errors for the corresponding pairwise error event.

Let us upper bound the UPEP of (\ref{UPEP}) using $\theta = \pi/2 $. Then, the diversity order of the STCM scheme can be calculated as
\begin{equation}
\xi =- \lim\limits_{E_b/N_0 \rightarrow \infty} \frac{\log P\big(\mathbf{Z} \rightarrow \mathbf{\hat{Z}}  \big) }{\log (E_b/N_0)} = RD
\end{equation}
which is also in accordance with the rank criterion for STBCs assuming that $ \big(\mathbf{Z} \rightarrow \mathbf{\hat{Z}}\big) $ is the worst case error event\cite{Jafarkhani}.

\textit{Remark}: Considering all possible realizations of $\mathbf{Z}$ and $\mathbf{\hat{Z}}$ for $M$-QAM/PSK constellations, the minimum value of $D$ is calculated as $D_{\textrm{min}}=2$ for Schemes 2 and 3, while $D_{\textrm{min}}=1$ for Scheme 1 due to their special system parameters given in (\ref{eq:8}). In the following, we describe the critical cases that affect $D_{\textrm{min}}$. For Scheme 1, when $\hat{k}=k$ \textit{and} $ \hat{l}=l $, its pairwise error events resemble those of Alamouti's STBC and a transmit diversity order of two is obtained. We also observe that $D=2$ for $\hat{k}\neq k$ \textit{and} $\hat{l}\neq l$ due to the structure of Scheme 1. On the other hand, for $ \hat{k}=k $ \textit{and} $ \hat{l} \neq l $ or $ \hat{k} \neq k $ \textit{and} $ \hat{l} = l $ along with $\hat{x}_1=x_1$ \textit{and} $\hat{x}_2=x_2$, transmit diversity orders provided by the corresponding pairwise error events reduce to unity. For this reason, Scheme 1 cannot provide transmit diversity. Scheme 2 always ensures $D=2$ due to its more regular structure. For $\hat{k}=k$, pairwise error events of Scheme 2 also resemble those of Alamouti's STBC, while for $\hat{k} \neq k$, a transmit diversity order of two is still retained. Finally, for Scheme 3, due to the clever selection of the system parameters (cross encoding of the active channel states) and thanks to the orthogonality of (\ref{STBC}), a transmit diversity order of two is obtained for all different cases mentioned above, i.e., for all possible $k,l,\hat{k},\hat{l} \in \left\lbrace 1, 2,\ldots,2^{M}\right\rbrace $ and $x_1,x_2,\hat{x}_1,\hat{x}_2 \in \mathcal{S}$.

\vspace*{-0.0cm}
\section{Simulation Results and Comparisons}
In this section, theoretical and computer simulation results are presented for the proposed STCM schemes and BER comparisons are performed with the reference schemes. We consider Gray and natural mappings for $Q$-QAM/PSK symbols and channel states, respectively. In Table II, we compare STCM schemes with Alamouti's STBC, STBC-SM, MBM-SIMO and MBM-MIMO $(T=2)$ schemes in terms of data rate $(\eta)$, transmit diversity order $(D_{\textrm{min}})$ and ML decoding complexity (total number of required metric calculations), where $C =  \lfloor{ \log_2 \binom{T}{2} } \rfloor$. As seen from Table II, STCM schemes provide interesting trade-offs among the considered system paramaters. For instance, Scheme 1 provides a high data rate and moderate decoding complexity; however, with a low diversity order. On the other hand, Scheme 2 is a compromise between Schemes 1 and 3, with moderate data rate and ML decoding complexity. Finally, Scheme 3 outperforms Schemes 1 and 2 in terms of diversity order and data rate, respectively; nevertheless, it has a high decoding complexity. 

\begin{table}[t]
	\centering
	\setlength{\extrarowheight}{1pt}
	\caption{Comparison of the proposed STCM and reference schemes \hspace*{0.3cm}
		(L: Low, M: Moderate, H: High)}
	\vspace*{-0.2cm}
	\begin{tabular}{cccc} \hline \hline
		Scheme & \begin{tabular}[c]{@{}c@{}}Data rate $(\eta)$\\ \end{tabular} & \begin{tabular}[c]{@{}c@{}}Tx. div. or.\\ $(D_{\text{min}})$ \end{tabular} & \begin{tabular}[c]{@{}c@{}}ML decoding\\ complexity \end{tabular} \\ 	 \midrule 
		Alamouti's STBC & $\log_2 Q $ (L) & $ 2 $ (H) & $ 2Q $ (L) \\
		STBC-SM\cite{STBC_SM} &  $0.5 C + \log_2 Q  $ (L)  & $2$ (H)  & $2^{ C +1}Q$ (M)  \\
		MBM-SIMO \cite{Khandani2} & $M+ \log_2 Q $ (H) & $ 1 $ (L) & $ 2^M Q $ (M) \\
		MBM-MIMO \cite{Khandani3} & $2M+ 2\log_2 Q $ (H) & $ 1 $ (L) & $ 2^{2M} Q^2 $ (H) \\
		STCM Scheme 1 & $ M + \log_2 Q $ (H) & $ 1 $ (L) & $ 2^{2M+1} Q $ (M) \\
		STCM Scheme 2 & $ 0.5M + \log_2 Q $ (M) & $ 2 $ (H) & $ 2^{M+1} Q $ (M) \\
		STCM Scheme 3 & $ M + \log_2 Q $ (H) & $ 2 $ (H) & $ 2^{2M} Q^2 $ (H) \\  \hline \hline
	\end{tabular}
	\vspace*{-0.4cm}
\end{table} 

In Fig. 4, we compare our computer simulation results with the theoretical ABEP curves obtained in Section IV for the proposed STCM schemes, where we considered $M=4$ and $\eta=5$ bpcu, with $R=1$ and $2$. As seen from Fig. 4, the diversity order of Scheme 1 is lower than those of Schemes 2 and 3, and the theoretical ABEP curves accurately predict the BER behavior for all schemes with increasing SNR values.

In Fig. 5, we investigate the BER performance of STCM schemes for $\eta=5$ bpcu and make comparisons with the reference systems. As seen from Fig. 5, the proposed STCM schemes provide considerable improvements in BER performance compared to MBM-SIMO scheme, which does not provide transmit diversity. Due to the information bits carried by the active channel states, the proposed STCM schemes can outperform the classical Alamouti's STBC, which requires higher order $M$-QAM constellations to reach a target data rate. 

In Fig. 6, we extend our computer simulations to $\eta=6$ bpcu transmission. We observe from Fig. 6 that the error performance difference between the classical Alamouti's STBC and the STCM schemes increases with increasing data rate. It is interesting to note that the proposed three STCM schemes provide an interesting trade-off between BER performance and decoding complexity. The ML decoding complexities of Schemes 1, 2 and 3 are calculated from Table II for $\eta=5$ bpcu as $1024$, $ 256 $ and $ 1024 $, respectively, while these values are equal to $ 2048 $, $ 512 $ and $ 4096 $ for $\eta=6$ bpcu. In all cases, we observe that Scheme 3 exhibits the best BER performance; however, its ML decoding complexity is considerably higher than those of Schemes 1 and 2. On the other hand, Scheme 2 has the lowest decoding complexity and outperforms Scheme 1 for the case of $R=2$. 

Finally, as seen from Figs. 5-6, the proposed STCM scheme also outperforms the STBC-SM scheme with $T=4$ and $8$ by using less number of transmit antennas since it benefits more efficiently from IM by exploiting the indices of the active channel states in data transmission. However, the price paid for this improvement is the increased decoding complexity.

\begin{figure}[!t]
	\begin{center}\resizebox*{\columnwidth}{7cm}{\includegraphics{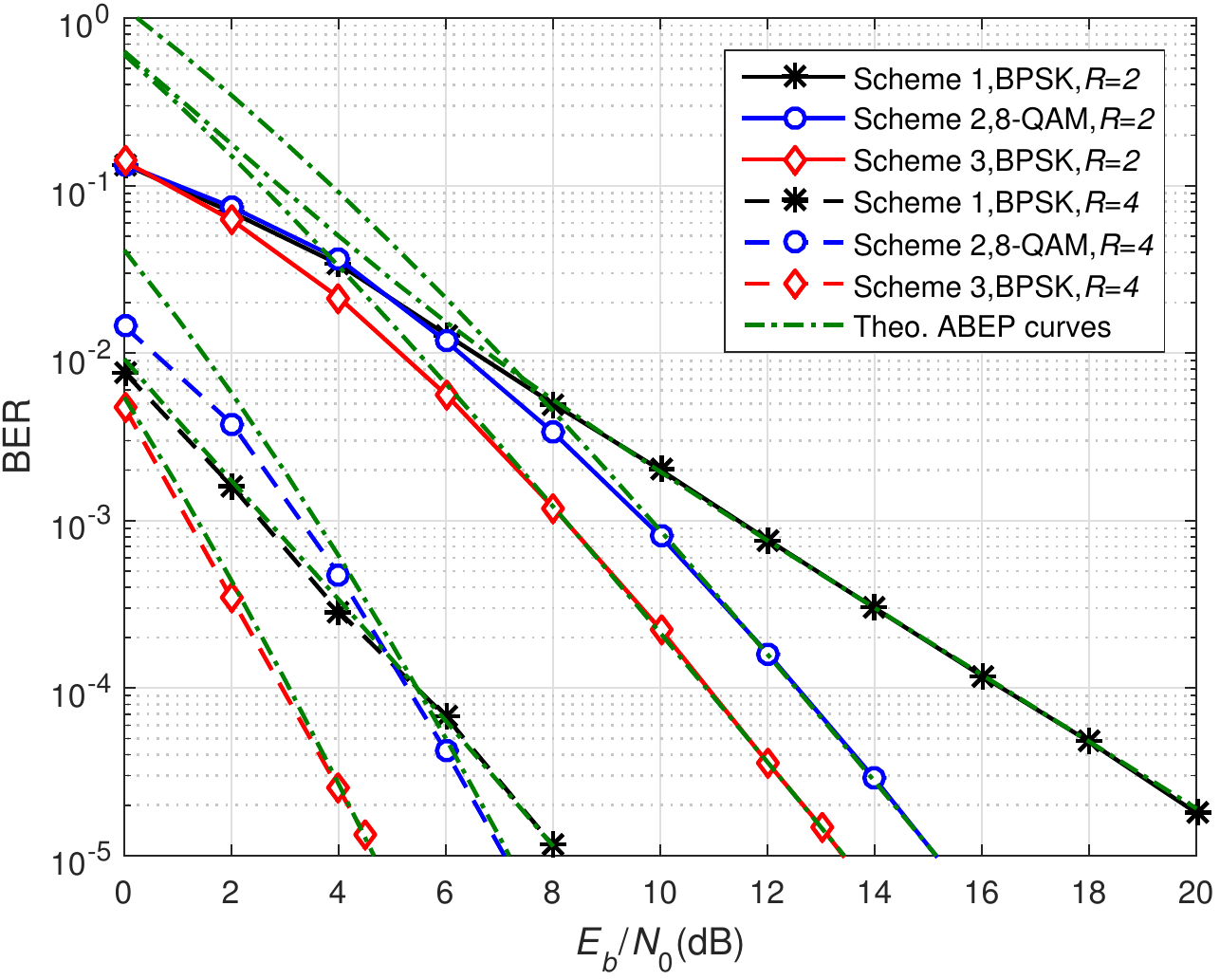}}
		\vspace*{-0.6cm}\caption{Comparison of theoretical ABEP curves with Monte Carlo simulation results for STCM schemes, $M=4$, $\eta=5$ bpcu, $R=2$ and $4$. }\vspace*{-0.3cm}
	\end{center}
\end{figure}

\begin{figure}[!t]
	\begin{center}\resizebox*{\columnwidth}{7cm}{\includegraphics{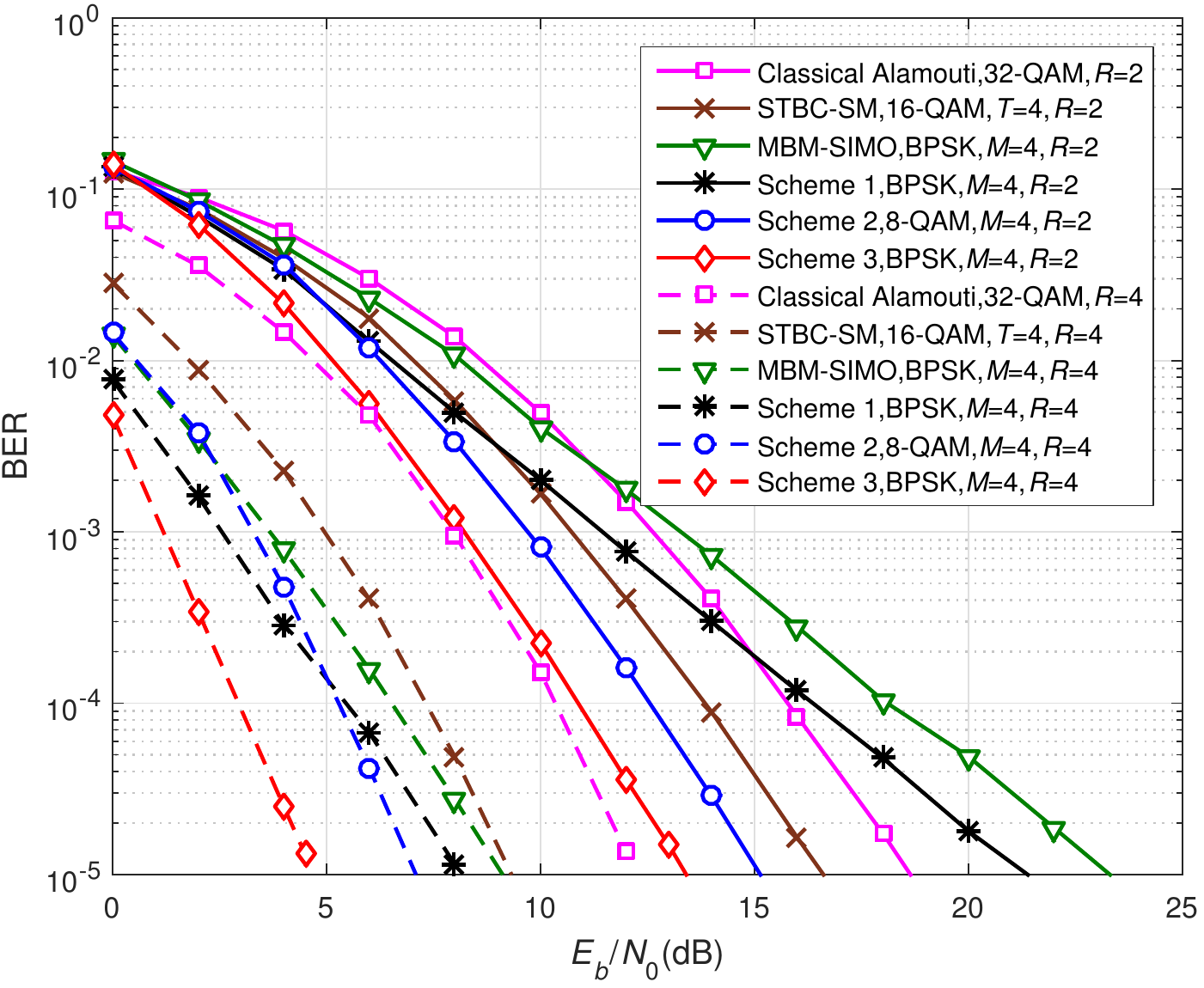}}
		\vspace*{-0.6cm}\caption{BER performance comparisons for $\eta=5$ bpcu, $R=2$ and $4$. }\vspace*{-0.3cm}
	\end{center}
\end{figure}

\vspace*{-0.0cm}
\section{Conclusions and Future Work}
In this paper, we have proposed the concept of STCM, which exploits space, time and channel state domains for the transmission of complex data symbols. The proposed STCM scheme can be considered as the generalization of either classical STBCs to channel state domain or plain MBM into space and time domains. It has been shown via computer simulations as well as theoretical ABEP calculations that the proposed STCM schemes can provide significant improvements in BER performance compared to classical Alamouti's STBC and plain MBM schemes. Several interesting research problems such as suboptimal detection methods, analyses for correlated channel states as well as MIMO fading channels, enhancement and generalization of the STCM scheme and the combination of STCM with space modulation techniques remain to be investigated. 

\begin{figure}[!t]
	\begin{center}\resizebox*{\columnwidth}{7cm}{\includegraphics{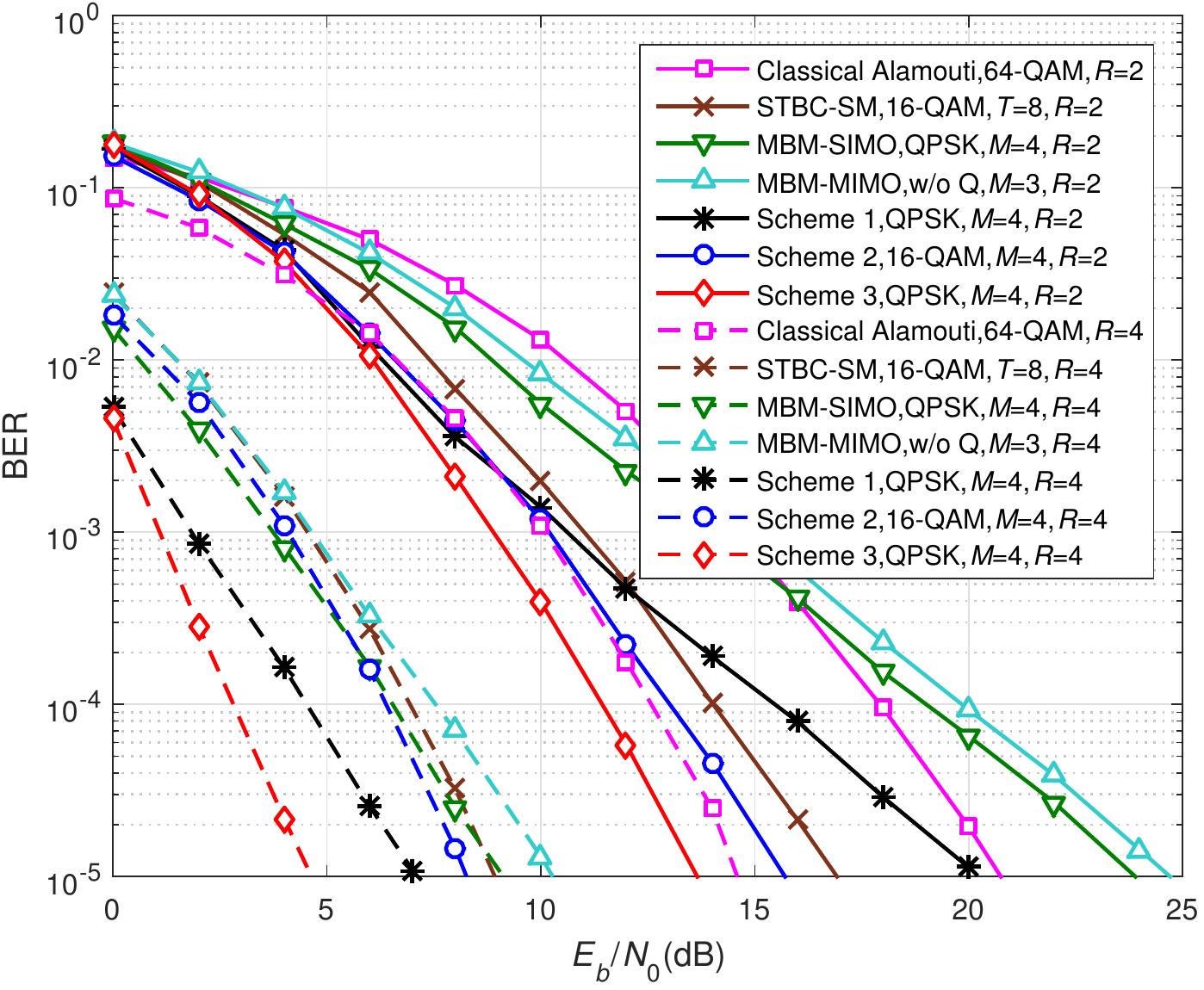}}
		\vspace*{-0.6cm}\caption{BER performance comparisons for $\eta=6$ bpcu, $R=2$ and $4$.}\vspace*{-0.3cm}
	\end{center}
\end{figure}

\vspace*{-0.5cm}
\bibliographystyle{IEEEtran}
\bibliography{IEEEabrv,bib_2017}
%

\end{document}